\documentclass[pra,twocolumn,amsfonts]{revtex4}
\usepackage{graphicx}
\usepackage{amsmath}
\usepackage{amsfonts}
\usepackage[dvips]{epsfig}
\usepackage{bm}
\usepackage{srcltx}
\usepackage{epsf}

\begin{document}
\author{S. Marcovitch, B. Reznik}
\title{Entanglement of Solitons in the Frenkel-Kontorova Model}
\affiliation{ School of Physics and Astronomy,
Raymond and Beverly Sackler Faculty of Exact Sciences,
Tel-Aviv University, Tel-Aviv 69978, Israel.}
\begin{abstract}
We investigate entanglement of solitons in the
continuum-limit of the nonlinear Frenkel-Kontorova chain.
We find that the
entanglement of solitons manifests particle-like behavior as they are characterized
by localization of entanglement.
The von-Neumann entropy of solitons mixes critical with noncritical behaviors.
Inside the core of the soliton the logarithmic increase of the entropy is faster than
the universal increase of a critical field, whereas
outside the core the entropy decreases and saturates the constant value of the
corresponding massive noncritical field.
In addition, two solitons manifest long-range entanglement that
decreases with the separation of the solitons more slowly
than the universal decrease of the critical field.
Interestingly, in the noncritical regime of the Frenkel-Kontorova model,
entanglement can even increase with the separation of the solitons.
We show that most of the entanglement of the so-called internal modes of the solitons
is saturated by local degrees of freedom inside the core,
and therefore we suggest using the internal modes
as carriers of quantum information.
\end{abstract}
\maketitle
\date{\today}

\section{Introduction}

It is well known that nonlinear theories manifest a variety of
classical configurations such as solitons
and instantons and dynamical effects such as symmetry breaking,
confinement and phase transitions,
which can never be seen using an ordinary perturbative expansion.
Such exact classical configurations can provide
us with powerful insight into the nonperturbative
content of those theories
and may exhibit novel emergent particle-like behavior.
Classical configurations such as solitons are commonly
observed solutions of nonlinear relativistic or non-relativistic equations.
They are localized, finite energy solutions, which can travel
while retaining their shape.
These classical configurations also strongly
determine the structure of the
corresponding quantized theories.
For example, the Hilbert space of the well-known one-dimensional
$\phi^4$ and sine-Gordon field theory models are classified to
topologically inequivalent sectors,
which cannot be obtained by power expansion of the global vacuum state.
However, these sectors can be obtained by quantizing a field around
the nontrivial backgrounds of either the
$\phi^4$ kinks or sine-Gordon solitons \cite{rajaraman}.

It is interesting to ask whether nonperturbative effects,
such as collective particle-like behavior,
are manifested in the entanglement structure of nonlinear theories.
Much of the recent study of entanglement in multipartite systems has been carried out
in linear systems
such as harmonic and spin chains \cite{nielsen,werner}.
In such one-dimensional system
it has been found that entanglement is a measure of criticality
and manifests a universal scaling behavior \cite{black_hole_analogy,modewise2}.
The behavior of entanglement in nonlinear systems
may challenge some of the insights from the linear systems,
and shed new light regarding the nature of entanglement.
The study of entanglement in nonlinear systems
may also suggest new implementations of quantum information processes.

In this work we investigate entanglement
in a chain of nearest neighbors interacting oscillators
subjected to a periodic on-site external potential,
known as the Frenkel-Kontorova (FK) model \cite{FK,kivshar}.
This model possesses rich behavior and so we parameterize it
from two aspects: its weak/strong-coupling regime and critical/noncritical regime.
The continuum-limit solutions of the FK model are the well-known sine-Gordon solitons.
Although we deal with a finite discrete system, it still retains
most of the characteristics of the continuum
and therefore in the following we shall use the same
field-theory terminology (vacuum-sector, soliton-sector, etc.).

Roughly speaking, one can say that a soliton solution is one in which
the effective $(\rm{mass})^2$ of the quantum free-field around the classical background
decreases inside the core of the soliton and even becomes negative.
Outside the soliton the field behaves as a regular massive free theory.
Therefore, the field interpolates critical-like behavior
inside the soliton's core and noncritical behavior outside the core.

It can be expected, therefore, that the von-Neumann entropy
of a region which coincides with the soliton
will manifest a mixture of critical and noncritical behaviors, depending on its size.
This is the first result we provide.
We compute the entanglement in the single-soliton sector ground state.
We find that as long as the block is confined to the soliton's core,
entanglement increases logarithmically but faster than
the critical harmonic systems.
However, as we keep increasing the size of
the block so that it grows outside the soliton's kink,
the entanglement decreases and saturates the asymptotic
constant limit that corresponds to the noncritical massive system.
Therefore, the soliton manifests a particle-like localization of entanglement.

Next we study long-range effects of entanglement, that is
the entanglement between two spatially separated blocks,
and use the logarithmic negativity as a measure of entanglement.
We find that the entanglement distribution is
peaked around the location of the solitons.
We then study soliton-soliton entanglement as a function of their
separation and compare the results
with the corresponding behavior in the vacuum sector.
The latter is also equivalent to the ground
state of a linear harmonic chain with mass $m\sim1\sqrt{g}$, where
$g$ is the coupling strength.

In the two-soliton sector we see complex behavior of long-range entanglement.
We find that the entanglement between two solitons is larger than
the corresponding entanglement in the vacuum sector.
For relatively small separations of the solitons the entanglement 
decays with the separation more slowly than
that of the critical vacuum sector.
Moreover, in the noncritical limit
entanglement loses one of its most
fundamental properties:
it is no longer a monotonically decreasing function of the distance between the solitons!

Much of our discussion is devoted to the strong-coupling limit.
In the weak-coupling limit, where the kinks become impurities that are confined to
a few particles, we find that the entanglement loses both its localization and long-range effects.

Finally, we suggest and investigate the possible use of
solitons as carriers of quantum information.
We propose utilizing the internal localized modes,
which describe collective and vibrational perturbations of the classical soliton,
as the carriers of quantum information.
The soliton is handled as a non-perturbative classical object
while the internal modes are quantized linear perturbations "around" it.
Since the internal modes are "attached" to the soliton,
by moving the soliton, quantum information can be transported in the system. 
(Transportation of quantum information has been discussed for 
harmonic chains \cite{remark+plenio}, spin chains
\cite{sc} and Josephson arrays \cite{ja}).
We also formulate multi-soliton processes,
which generate entanglement.

The paper is organized as follows.
In the next section we describe the FK model and analyze its
strong/weak-coupling and critical/noncritical limits.
Then in section $3$ we investigate localization of entanglement
in the continuum limit of the FK model.
In section $4$ we present the soliton-soliton long-range entanglement.
In section $5$ we briefly discuss the loss of both long-range effects and
localization in the weak-coupling limit.
In section $6$ we discuss possible quantum information implementations.
Lastly, in section $7$ we
summarize the significant results and outline future directions of research.

\section{Frenkel-Kontorova model}
\begin{figure}[ht]
\center{
\includegraphics[width=3in]{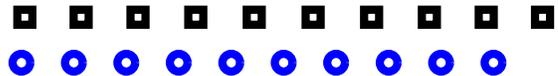}
\caption{(color online). Schematic structure of the FK chain for $N=10$. The blue circles
are the vacuum equilibrium locations, where the black squares represent
the soliton solution.
}
\label{fig0}
}
\end{figure}

We consider $N$ particles with canonical coordinates $\phi_n$ and $\pi_n$,
described by the Hamiltonian of the FK model \cite{FK,kivshar}:
\begin{equation}
H=\sum_{n=1}^N\frac{1}{2}(\pi_n^2+2V_{sub}(\phi_n)+g(\phi_{n+1}-\phi_n)^2),
\label{fk}
\end{equation}
where $\phi_n$ is the displacement of the $n$th particle from its equilibrium position,
$g$ is the coupling constant between particles in the chain
and
\begin{equation}
V_{sub}(\phi)=1-\cos\phi.
\label{vfk}
\end{equation}
We use dimensionless units in which the particles mass equals $1$ and
the period and amplitude of the substrate potential $V_{sub}(\phi)$
are $a_s=2\pi$ and $\epsilon_s=2$ respectively.
In the ground state of the system all particles occupy minima of the potential
such that $\phi_n=0$ or $\phi_n=2\pi$, etc.
Then the simplest excited state is a soliton that connects two neighboring ground state solutions as illustrated in figure \ref{fig0}.

\subsection{Classical solutions}
The exact solutions of the discrete FK chain are not known in an explicit analytical form
due to the discreteness of the problem.
However, in the strong coupling limit, $g>>1$,
an approximate discrete solution
can be obtained using the {\it{continuum limit approximation}},
$n\rightarrow x=n a_s$ and $\phi_l(t)\rightarrow\phi(x,t)$
\cite{philip}.
In this framework we begin with the sine-Gordon Hamiltonian density
\begin{equation}
\label{lagrangian}
\mathcal{H}(x,t)=\frac{1}{2}\pi^2+g a_s^2\frac{1}{2}(\phi')^2+V_{sub}(\phi),
\end{equation}
where a dot or a prime represents differentiation with respect to
time or space respectively.
The equation of motion is then the integrable sine-Gordon (SG)
equation:
\begin{equation}
\label{sg}
\frac{\partial^2\phi}{\partial t^2}-g a_s^2\frac{\partial^2\phi}{\partial x^2}+\sin\phi=0.
\end{equation}

Imposing the boundary conditions 
\begin{equation}
\frac{d\phi}{dx}(\pm\infty)=0
\label{boundarysg}
\end{equation}
and rescaling $x\to x/a_s$,
we obtain the SG soliton solution:
\begin{equation}
\phi_{SG}(x)=4\tan^{-1}\exp\Big(-\sigma\frac{x-X}{\sqrt{g}}\Big),
\label{sgkink}
\end{equation}
where $\sqrt{g}$ defines the soliton's width in units of the period of the substrate potential,
$\sigma=\pm1$ and $X$ is the coordinate of the soliton's center.
For $g>g_{con}\sim 16$ the continuum limit is a reasonable approximation.
We can then describe our $N$-particle system using the continuum solution
by sampling $\phi_{SG}$ at $N$ points with separation $1$,
where $X$ is sampled at the middle.
Throughout the paper we shall refer to this sampled classical solution (Eq. \ref{sgkink})
simply as the soliton solution (for $g>g_{con}$).
Note that as we increase $g$,
even though the separation between particles is kept constant,
the number of particles (points) 
that sample the core of the soliton, $\sqrt{g}$, increases.

\begin{figure}[ht]
\center{
\includegraphics[width=3in]{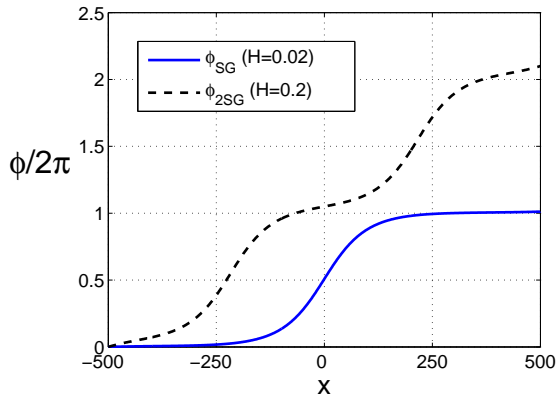}
\caption{(color online). Single soliton $\phi_{SG}$ and double-soliton $\phi_{2SG}$ solutions of the finite
SG equation sampled at $N=1000$ points.
The length of the region is given by $2L=16\lambda_J$ where $\lambda_J$ is the Josephson length.
For the single soliton solution the stability 
region for the magnetic field is $H\in[0.0019,1.0052)$
and for the double-soliton $H\in[0.1023,1.0622)$.
See details in Appendix A.
}
\label{fig1}
}
\end{figure}

In order to investigate long-range entanglement between solitons,
we are interested in exploring
double-soliton solutions as well.
There are no multi-soliton static solutions in infinite systems.
However, the finite sine-Gordon equation also possesses multi-soliton
static solutions (Eq. \ref{sg}).
This equation describes the magnetic flux in a
long Josephson junction of length $2L$ with a constant
homogeneous external magnetic field $H$, perpendicular to the barrier
and with bias current line density $I$.
We use these double-soliton solutions, $\phi_{2SG}$,
as the continuum limit approximation of the of FK model
in a finite system.
In appendix A we describe the analytical solutions of the double-soliton solutions.
For illustration purposes, we present in figure \ref{fig1} single soliton and double-soliton 
configurations.
More details are provided in Appendix A.

In the weak-coupling limit of the FK model ($g<g_{con}$) there are no known analytical solutions.
Still, in this regime there are kink solutions, $\phi_{kink}$,
which are localized to a small number of particles.
These kink solutions can be found numerically.
We use a descending gradient
algorithm that minimizes the classical configuration
energy, given an initial solution to start with.
Apparently, in the weak-coupling limit, where $g$ is small ($g\sim1$)
the analytical continuum term is still a good enough starter
for the minimizer to find the minimal stable configuration.

\subsection{Quantization}
In order to explore quantum mechanical effects in solitons we use a semi-classical framework.
For simplicity, we provide the analysis on the continuous solutions.
We consider small perturbations $\eta(x,t)$ on the background of
the classical solution $\phi_0(x)$ (single soliton, double-soliton or kink):
$\phi(x)=\phi_0(x)+\eta(x,t)$ \cite{rajaraman}.
Setting $H=U+T$,
where $U$ is the potential energy and $T$ is the kinetic energy,
we assume the solutions are static, $T=0$.
Then we can make a functional Taylor expansion of the potential energy
$$U(\phi)\equiv \int dx\Big[\frac{g}{2}(\phi')^2+V_{sub}(\phi)\Big]$$ about $\phi_0$:
\begin{equation}
\label{variational}
U\!(\!\phi\!)\!=\!U\!(\!\phi_0\!)\!+\!\int\! dx\! \frac{1}{2}\!\Big\{\!\eta(x)\!\Big[\!-g\nabla^2+\Big(\!\frac{\partial^2 V_{sub}}{\partial \phi^2}\!\Big)\!_{\phi_0}\!\Big]\!\eta(x)\!\Big\},
\end{equation}
where cubic and higher terms in $\eta$ are neglected.
This requires that the magnitude of $\eta(x)$ as well as
the third and higher derivatives of $V(\phi)$ at $\phi_0$ be small.

The eigenvalues and eigenfunctions of the operator 
$(-g\nabla^2+\partial^2 V_{sub}/\partial \phi^2)$
evaluated at $\phi(x)=\phi_0(x)$ are then the generalized solutions of the Schrodinger-like
equation
\begin{equation}
\label{stability}
\Big[-g\nabla^2+\Big(\frac{\partial^2 V_{sub}}{\partial \phi^2}\Big)_{\phi_0(x)}\Big]\eta_l(x)=\omega_l^2\eta_l(x),
\end{equation}
where the $\eta_l(x)$ are the orthonormal normal-modes of the fluctuations around
$\phi_0(x)$.
In the continuous solution $l$ is a continuous index with possible additional discrete values.

Translating the above analysis to the discrete system we obtain 
a set of equations,
that is, the $\eta_l(n)$ and $\omega_l^2$ are obtained by
diagonalizing the matrix $B(n,n)=2g+\cos\phi_0(n)$, $B(n,n\pm 1)=-g$.
We require that $w_l^2>0$; otherwise the system is in a nonstable configuration.

\begin{figure}[ht]
\center{
\includegraphics[width=3in]{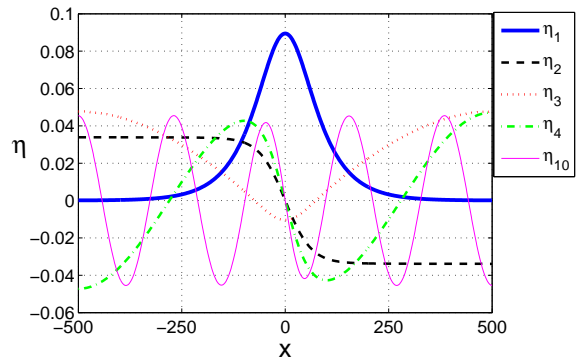}
\caption{(color online). Normal modes in the background of the single soliton solution,
shown in figure \ref{fig1}.
The four first normal modes and the $10$'th
normal mode are shown. The first normal mode is an internal mode.
}
\label{fig2}
}
\end{figure}
\begin{figure}[ht]
\center{
\includegraphics[width=3in]{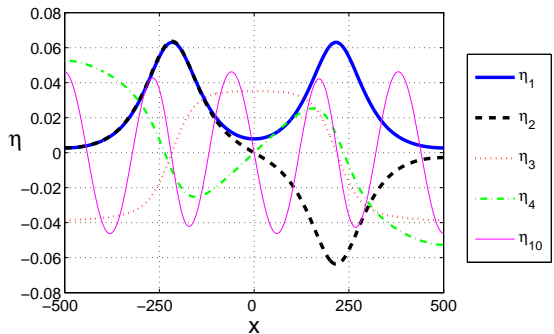}
\caption{(color online). Normal modes in the background of the double-soliton solution
shown in figure \ref{fig1}.
The four first normal modes and the tenth
normal mode are shown. The first two normal modes are internal modes.
}
\label{fig3}
}
\end{figure}

We obtain the quantum-mechanical modes of the system by setting:
\begin{equation}
\label{eta}
\eta(n,t)=\sum_l c_l(t)\eta_l(n),
\end{equation}
and quantizing the normal-mode coefficients $c_l$.
The canonical degrees of freedom are then:
\begin{equation}
\label{fields}
\begin{split}
&\phi(n)=\phi_0(n)+\sum_l\eta_l(n)\frac{1}{\sqrt{2\omega_l}}(a_l e^{i\omega_l t}+a_l^\dagger e^{-\i \omega_l t}),
\\& \pi(n)=\sum_l-i\eta_l(n)\sqrt{\frac{\omega_l}{2}}(a_l e^{i\omega_l t}-a_l^\dagger e^{-i \omega_l t}),
\end{split}
\end{equation}
where $a_l$ and $a_l^\dagger$ are annihilation and creation operators of the normal modes, and
\begin{equation}
\label{spectrum}
E_{\{n_l\}}=U(\phi_0)+\sum_l\Big(n_l+\frac{1}{2}\Big)\omega_l
\end{equation}
where $n_l$, is the excitation number of the $l$'th normal mode.

The background classical solution $\phi_0$ could
either be an absolute minimum or a local minimum of $U(\phi)$.
Consider first the case that $\phi_0$ is an
absolute minimum.
Then all particles are
located at their equilibrium positions 
and the solution does not contain kinks or solitons.
The normals modes are then
identical with the "phonon modes" of a harmonic chain:
\begin{equation}
\label{etavacuum}
\eta_l(n)\propto\exp(ik_l n);\quad k_l=2\pi l/N
\end{equation}
and the spectrum is given by \cite{remark_omega}:
\begin{equation}
\label{vacuumspec}
\omega_l^2=1+2g(1-\cos k_l).
\end{equation}

Upon quantization, the quantum states of the FK system in the absolute minimum case,
which we also refer to as the "vacuum sector",
coincide with that of a harmonic chain.
Hence, in the continuum limit the vacuum sector of the FK model
corresponds to a free massive scalar field with $m\sim 1/\sqrt{g}$.
and the ground state has the same properties of the corresponding free field vacuum state.

If $\phi_0$ is a local minimum, then the solution contains solitons
(or kinks).
In the continuum limit the eigenvalue equation is
\begin{equation}
\Big[-g\frac{\partial^2}{\partial x^2}+\cos\phi_{SG}\Big]\eta_l(x)=\omega_l^2 \eta_l(x).
\end{equation}
In the limit of an infinite system the translational invariance of the system
gives rise to a zero frequency mode (Goldstone mode, or "zero mode"),
which requires special attention.
This problem is avoided in our case due to discreteness and finiteness of the system.
Nevertheless, in correspondence with the infinite limit,
the zero mode is still characterized
by a localized shape of a bound state and will be referred to as an
{\em internal mode}.
(In other nonlinear models which have soliton
solutions such as the $\phi^4$ model, there can be additional internal
localized modes which are not zero modes.)

For clarification, 
we present several normal modes and their eigenfrequencies for the solutions presented
in figure \ref{fig1}.
In figure \ref{fig2} the first $4$ normal modes and the $10$'th normal mode
of the single-soliton solution are shown, where the first $4$ eigenfrequencies are
$\omega_1\simeq0.0007$, $\omega_2\simeq1.0002$, $\omega_3\simeq1.05$, $\omega_4\simeq1.2$,
given in dimensionless units.
Note that the first eigenvalue is less than $1$ and corresponds to the internal mode,
while all the others are phonon modes.
In figure \ref{fig3}, the first $4$ normal modes and the $10$'th normal mode
of the double-soliton solution are shown,
where the first $5$ eigenfrequencies are $\omega_1\simeq0.008$, $\omega_2\simeq0.016$, $\omega_3\simeq1.013$,
$\omega_4\simeq1.07$, $\omega_5\simeq1.3$.
Here the first two normal modes are internal modes.

As we shall see, it is the presence of the internal modes
which gives rise to qualitatively different behavior of the quantum mechanical states in
the solitonic sector,
such as particle-like behavior and localization of
entanglement.

\subsection{Correlations in the vacuum and soliton sectors}

We would like to compare the behavior of correlations in the ground states
of the vacuum and soliton sectors in the strong and weak coupling regimes.
Let us define the correlation lengths $\xi_n=\langle\phi_0\phi_n\rangle$
and $
\nu_n=\langle\pi_0\pi_n\rangle$,
where $n$ is the (integer) distance between the particles.

We begin with the strong-coupling regime.
In the vacuum sector the correlation lengths $\xi_n,\nu_n$ have been thoroughly
analyzed in connection with the harmonic chain \cite{modewise2}.
These correlations are classified by the critical and noncritical regimes.
In the critical limit there are long-range correlations
as there is no length scale in the system ($m\to0$).
In this limit the correlations scale as:
\begin{equation}
\begin{split}
\xi_n\sim &\log(\frac{1}{n}),
\\\nu_n\sim &-\frac{1}{n^2}.
\end{split}
\end{equation}
In a system with $N$ particles and a coupling strength $g$
the critical regime is valid if $N<<\sqrt{g}$.

The noncritical regime is characterized by a finite length scale.
Here the correlations decay exponentially with the distance:
\begin{equation}
\begin{split}
\xi_n&\sim-\frac{e^{-n}}{n^{1/2}},
\\\nu_n&\sim\frac{e^{-n}}{n^{3/2}}.
\end{split}
\end{equation}
The noncritical regime is defined for $N>>\sqrt{g}$.
For $N=1000$ \cite{remark_N}, for example, the system is critical for $g>10^7$,
and noncritical for $g_{con}<g<10^5$ \cite{remark_alpha}.

We turn now to the soliton sector.
For the critical limit the number
of particles in the core of the soliton
should be greater than the number of particles in the whole chain:
$N<\sqrt{g}$.
Therefore, in the critical limit the vacuum sector and the soliton sector coincide:
for a constant $N$ taking $g$ to infinity causes the loss of the internal mode,
which now enters into the phonon band.
We define $g_{max}(N)$ as the maximal coupling constant
for which there is a discrete internal mode.
For $g>g_{max}(N)$ no difference is predicted
between the ground states of the vacuum and soliton sectors
since their spectra and normal modes become practically identical.

On the other hand, in the weak-coupling limit, $g<g_{con}$,
the correlations decay very fast and again one may expect no significant
difference between the vacuum sector and the kink sector.
Therefore, we expect qualitative difference between the vacuum sector and
the soliton sector only in the noncritical, strong coupling regime.

\begin{figure}[ht]
\center{
\includegraphics[width=3in]{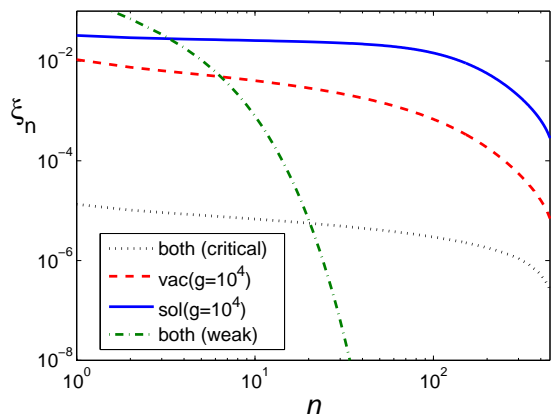}
\caption{(color online). Log-log scale of $\xi_n$, where $0\leq n\leq500$.
The plots are given for the vacuum sector and the soliton sector in the critical
($g=10^{10}$), noncritical ($g=10^4$) and weak-coupling ($g=5$) regimes.
In the soliton sector the center of the soliton is located at $n=0$.
}
\label{xx_cor}
}
\end{figure}

In figure \ref{xx_cor} we show on log-log scale $\xi_n$
for $N=1000$, where $0\leq n\leq500$.
The plots are given for the vacuum sector and the soliton/kink sector of the FK model,
where the center of the soliton/kink is located at $n=0$.
We study the following coupling constants:
$g=10^{10}$ (critical limit),
$g=10^4$ (noncritical regime)
and $g=5$ (weak-coupling limit).
First of all, we observe that the vacuum and soliton/kink sectors
are indeed undistinguishable both in the critical limit and the weak-coupling limit.
In the noncritical regime, however, $\xi_n^{sol}$ is qualitatively different
from $\xi_n^{vac}$.
Inside the soliton's core ($n<\sqrt{g}=100$),
$\xi_n^{sol}$ decays more slowly than $\xi_n^{vac}$ and even $\xi_n^{critical}$.
As $n$ becomes greater than $\sqrt{g}$, however, we see a strong decay in
$\xi_n^{sol}$, which quickly
becomes parallel to the graph of $\xi_n^{vac}$ in the noncritical regime.

We explain this behavior on physical grounds by relating to the effective
$(\rm{mass})^2$ of the linearized free-field around the critical vacuum and the
(noncritical) soliton sector:
\begin{equation}
\begin{split}
m&_{critical}^2\to0,
\\m&_{sol}^2(x)\sim\cos(\phi_0(x))=\left\{ \begin{array}{ll}
-\frac{1}{\sqrt{g}}, & x=0\\
\frac{1}{\sqrt{g}}, & |x|>>\sqrt{g}\\
\end{array} \right.
\end{split}
\end{equation}
As  $(\rm{mass})^2$ become negative for $x<\sqrt{g}$
the correlation length inside the core decays more slowly than that of the
critical vacuum sector.
Outside the core the correlation restores the massive vacuum sector behavior.
We expect that the entanglement of solitons will have a similar behavior
to the correlation length.

\section{Localization of soliton entanglement}
We turn now to the entanglement of the
FK model in the strong-coupling limit.
Both in the vacuum sector and in the soliton sector we assume the normal modes
are in their ground states, therefore the quantum-mechanical states are Gaussian.
Following \cite{modewise2}, let us represent the local canonical
variables of our N-mode system (Eq. \ref{fields})
by the vector
$$y=(\phi,\pi)^T,$$
where $\phi=(\phi_1,\phi_2,\dots,\phi_N)$ and $\pi=(\pi_1,\pi_2,\dots,\pi_N)$.
The commutation relations may thus be expressed as
$$[y_\alpha,y_\beta]=iJ_{\alpha\beta},$$
where J is the so-called symplectic matrix:
\[ J=\left( \begin{array}{cc}
0 & 1\\
-1 & 0 \end{array} \right) \]
Assuming $\langle y \rangle=0$,
the state of the system is entirely characterized by the matrix of the second moments,
the so-called phase-space $2N\times 2N$ covariance matrix (CM):
$$M(y)=\rm{Re}\langle y y^T \rangle.$$
Under a symplectic transformation $\tilde{y}=Sy$
a Gaussian state $M$ is mapped into a Gaussian state  $\tilde M=SMS^T$,
where $SJS^T=J$.
A theorem due to Williamson \cite{williamson, simon1}
states that there always exists a certain symplectic transformation
$S_W$ that brings $M$ to the normal form ("Williamson form"):
$$W=S_WMS_W^T=\rm{diag}(\lambda_1,\lambda_2,\dots,\lambda_N,\lambda_1,\lambda_2,\dots,\lambda_N),$$
where the diagonal elements $\lambda_1,\lambda_2,\dots,\lambda_N$
are referred as the symplectic eigenvalues
and must be greater or equal to $1/2$
according to the uncertainty principle.

Now, suppose our pure $N$-mode system (Eq. \ref{fields})
is partitioned into two sets $y_A$ and $y_B$.
In order to measure the entanglement between parts $A$ and $B$
we bring the reduced covariance matrices $M_A(y_A)$ and $M_B(y_B)$
into their Williamson normal forms.
Both $M_A(y_A)$ and $M_B(y_B)$ have the same $\lambda_j>1/2$ symplectic eigenvalues.
The entanglement is then measured by the von-Neumann entropy $E_S$:
\begin{equation}
\label{ent}
E_S=\sum_{\lambda_j}S(\lambda_j),
\end{equation}
where
\begin{equation}
\label{lamda}
S(\lambda)=\left(\lambda+\frac{1}{2}\right)\ln\left(\lambda+\frac{1}{2}\right)-
\left(\lambda-\frac{1}{2}\right)\ln\left(\lambda-\frac{1}{2}\right).
\end{equation}
Since the $\phi-\pi$ correlations vanish in our model,
$\lambda_j$ are given
by the square roots of the eigenvalues of $H_AG_A$ (or $G_AH_A$) \cite{modewise2}, where
\begin{equation}
\label{gh}
G=\langle\phi\phi^T\rangle,\quad H=\langle\pi\pi^T\rangle.
\end{equation}

\subsection{Mixing critical and noncritical behaviors}
\begin{figure}[ht]
\center{
\includegraphics[width=3in]{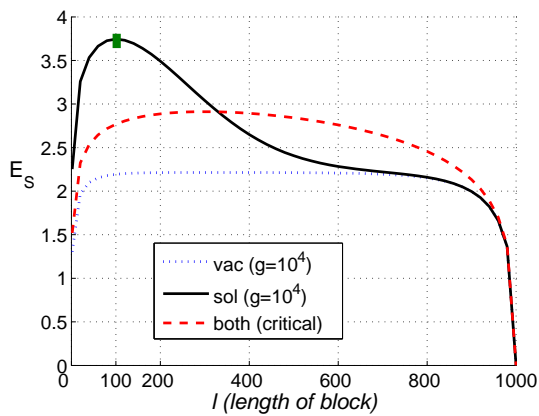}
\caption{(color online). Localization of entanglement in the soliton sector of the FK model.
In the noncritical regime ($g=10^4$) the size of the core is $\sqrt{g}=100$.
The critical limit is given for $g=10^8$.
Localization of entanglement is seen in the noncritical regime,
where maximal entropy is obtained for $l=\sqrt{g}$.
}
\label{vn_scale}
}
\end{figure}

\begin{figure}[ht]
\center{
\includegraphics[width=3in]{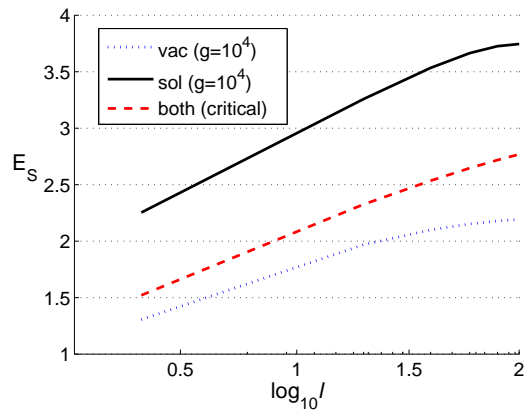}
\caption{(color online). Localization of entanglement in the soliton sector of the FK model, logarithmic scale.
The increase of entanglement inside the soliton is faster than the corresponding
one in the vacuum sector. Both entropies coincide outside the soliton.
The entanglement of the soliton sector has a maximal value and is therefore localized.
Note that the logarithmic increase inside the soliton is even faster than
the corresponding one in the {\it{critical}} vacuum sector.
}
\label{vn_scale_2}
}
\end{figure}
In figure \ref{vn_scale} we plot the von-Neumann entropy
of a block of $l$ particles in the chain
where in the soliton sector the center of the soliton coincide with the center of the block.
The plots are given for both the vacuum sector and the soliton sector
(in the strong-coupling regime of the FK model).
The plots are given for two regimes: critical ($g=10^8$),
in which case both sectors have the same behavior and noncritical ($g=10^4$).

First of all, let us examine the behavior in the vacuum sector.
In the critical regime the entanglement increases as the size of the block increases.
This effects stops, however, due to the finiteness of the chain
(due to edge effects this happens long before $l=N/2$).
In the noncritical regime
the entanglement increases for a finite region and then saturates a constant value.

In figure \ref{vn_scale_2} the same graph is plotted on logarithmic scale for
$l\leq100$.
We see that the critical vacuum sector fits the linear curve for $l<10^{1.5}\sim 30$,
where the noncritical vacuum sector fits the linear curve for $l<16$.

In the soliton sector we observe 
in figure \ref{vn_scale}
qualitatively different behavior,
as anticipated from the discussion on the correlation length in the previous section.
The von-Neumann entropy of the soliton {\it{mixes both the critical and noncritical scalings}}:
it scales logarithmically inside the soliton,
that is for $l< \sqrt{g}=100$,
and then it decreases to the noncritical vacuum sector asymptotic value.
As seen in figure \ref{vn_scale_2},
inside the soliton the logarithmic increase of the soliton sector is faster than
the corresponding vacuum sector.
This characteristic behavior implies that the entropy in the soliton sector must have a maximal value.
In contrast to the vacuum sector, the entanglement in the soliton sector
is localized as it reaches a maximal value at a certain length scale corresponding
to the size of the soliton.
Note that the logarithmic increase of the entanglement
inside the core of the soliton is even faster than
that of the {\it{critical}} vacuum sector.
This result corresponds to the effective negative
$(\rm{mass})^2$ inside the core.

\begin{figure}[ht]
\center{
\includegraphics[width=3in]{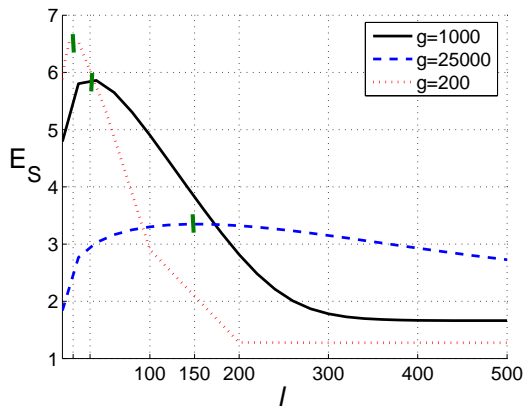}
\caption{(color online). Localization of entanglement in the soliton sector for $g=200,1000,25000$.
Entanglement localization "measures" the size of the soliton,
as maximal entanglement is achieved for $l=\sqrt{g}$ in all three cases.
}
\label{vn_scale_3}
}
\end{figure}

For reference we plot in figure \ref{vn_scale_3} the
von-Neumann entropy of the soliton sector for $g=200$, $g=1000$ and $g=25000$.
We may say that entanglement localization "measures" the size of the soliton,
as maximal entanglement is achieved for $l=\sqrt{g}$ in all three cases.
We can see that as the system enters deep into the noncritical regime ($g=1000$, $200$)
the entanglement becomes more localized
and reaches a higher maximum.

\subsection{Logarithmic prefactor}
\begin{figure}[ht]
\center{
\includegraphics[width=3in]{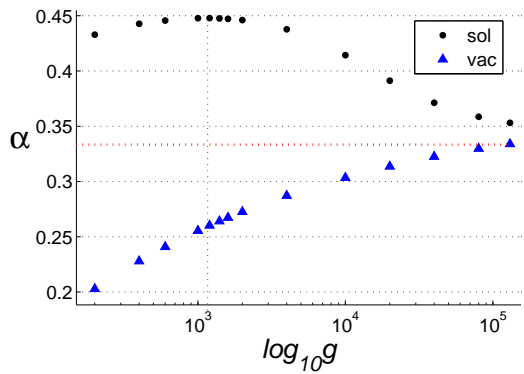}
\caption{(color online). The prefactor $\alpha(g)$ of $E_S=\alpha\log l$ for $l<<\sqrt{g}$.
Interestingly, for any value of $g$ in the strong-coupling regime ($g>g_{con}$),
$\alpha^{sol}(g)>\alpha^{vac}(g\to\infty)=1/3$.
In the soliton sector $\alpha_{max}^{sol}(g)\sim4/9$, and is obtained for $g=N$.}
\label{vn_max_g}
}
\end{figure}

We would like now to obtain the prefactor $\alpha(g)$ in the logarithmic scaling $E_S=\alpha(g)\log l$
for small enough values of $l$, $l<<\sqrt{g}$, in the vacuum sector and the soliton sector.
As a first order for the linear curve we take
the first two points in the graph: $(\log2,E_S(\log2))$ and $(\log4,E_S(\log4))$.
For $g\rightarrow\infty$ both models have approximately the well-known universal factor of $\alpha_c=1/3$
in critical bosonic one-dimensional fields.
In figure \ref{vn_max_g} we plot $\alpha(g)$ for
$g\leq g_{max}(N=1000)\sim1.3*10^5$, where $g$ is given in a $\log_{10}$ scale.
We observe that as $g$ gets closer to $N$ the initial logarithmic 
jump of the entanglement becomes sharper.
This is explained from the following conditions:
\begin{equation}
\begin{split}
\alpha^{sol}(g)>\alpha^{vac}(g),
&\\ E_S^{sol}(g)\underset{l>>\sqrt{g}}{\rightarrow} E_S^{vac}(g).
\end{split}
\end{equation}
Therefore, as $g$ decreases $E_S^{sol}$ falls more rapidly outside the core  
and the initial logarithmic jump of the entanglement becomes sharper.
This effect stops, however, for $g<N$, as then already for $l=1$, the entanglement
is close to its maximal value.
As anticipated from the correlations behavior,
for any value of $g$ in the strong-coupling regime ($g>g_{con}$),
the logarithmic increase is even faster than the corresponding one
in the critical vacuum sector:
$\alpha^{sol}(g)>\alpha^{vac}(g\to\infty)=\frac{1}{3}$, where
$\alpha_{max}^{sol}(g)\sim4/9$ is obtained for $g=N$.

\subsection{Maximal entropy}
\begin{figure}[ht]
\center{
\includegraphics[width=3in]{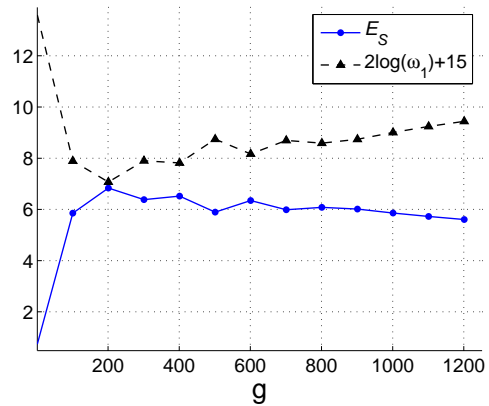}
\caption{(color online). Maximal entropy as a function of $g$.
$\omega_1$ is minimal for $g=200$, exactly where the von-Neumann entropy reaches its
maximal value.
}
\label{maximal_ent}
}
\end{figure}

Maximal localized entanglement is achieved in the soliton sector
for lower values of the coupling constant $g$
(yet not too low, so that the strong-coupling limit still holds).
We show that the maximal value of the entanglement 
is determined by the frequency of the internal model.
In figure \ref{maximal_ent}
the maximal localized von Neumann entropy as a function of $g$ is shown, where
for reference, we plot on the same graph $2\log(\omega_1)+15$,
so that both plots appear with the same scale.
Clearly, $\omega_1$ is minimal exactly where the von-Neumann entropy reaches its
maximal value ($g=200$).

In passing we note that 
as $g$ gets closer to the critical limit, 
$\omega_1$ increases and eventually
penetrates the phonon band and stops behaving as a bound state.
In addition, $\omega_1$ increases as discreteness increases ($g$ decreases).
This is associated with the existence of a potential barrier in a discrete lattice
that must be overcome to move the soliton
along one lattice spacing,
and is known as Peierls-Nabarro potential \cite{pn}.
Note that $\omega_1$ has an oscillatory behavior for $g>200$.
Oscillations of $\omega_1(g)$ are known for models that
deviate from the FK sinusoidal potential \cite{oscillate,kivshar2}.
Here we observe the oscillations in the FK model for large values of $g$,
where $\omega_1$ starts to increase.

\section{Long-range soliton-soliton entanglement}
We now explore long-range effects of entanglement in the FK model.
We are interested in the entanglement between two separated blocks of particles $A$ and $B$
with the same number of particles $l$, separated by $d$ particles.
Now $M_{AB}$ does not constitute a pure-state and the von-Neumann
entropy cannot be used anymore.
We therefore use a mixed-state entanglement measure -- the logarithmic negativity (LN) \cite{vidal,simon3}.
This measure is based on the following observation:
when the parts are entangled,
reversing time direction in one part of the system:
$y_{AB}^{PT}=(\phi_A,\phi_B,\pi_A,-\pi_B)^T$,
breaks the symplectic symmetry of their covariance matrix,
yielding $\lambda_j<1/2$, which violates the uncertainty principle.
The amount by which $\lambda_j<1/2$ is a measure of the entanglement:
\begin{equation}
\label{logneg}
E_{LN}=-\sum_j \ln (2\lambda_j).
\end{equation}

\subsection{Distribution of long-range entanglement}
\begin{figure}[ht]
\center{
\includegraphics[width=3in]{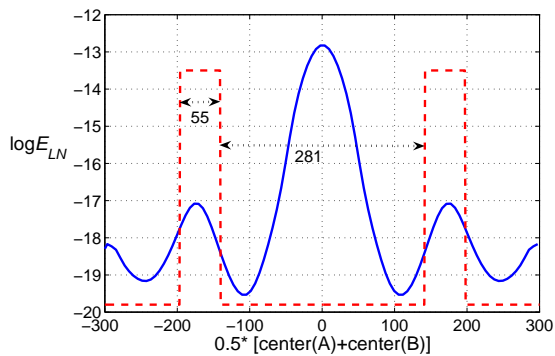}
\caption{(color online). Long range entanglement between two sliding blocks in the presence of two solitons
with $g=3000$, $l=l_{sol}=56$ and $d=d_{sol}=281$.
LN is maximized when the blocks coincide with the solitons' cores
(the average of the centers is located in the center of the chain, $n=0$).
}
\label{sliding}
}
\end{figure}

\begin{figure}[ht]
\center{
\includegraphics[width=3in]{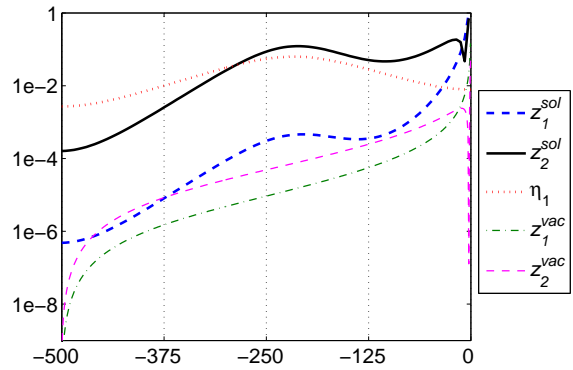}
\caption{(color online). First and second participation functions of the double-soliton
and vacuum sectors, shown in logarithmic scale,
where the chain is divided into two equal-sized blocks.
In contrast to the localized behavior of the participation functions
in the vacuum sector,
the participation functions in the solitons sector
peak at the core of the soliton.
}
\label{partic2}
}
\end{figure}

In the double-soliton sector of the FK model
we explore long-range entanglement between two solitons,
which are characterized by the length scale $l_{sol}\sim\sqrt{g}$.
We obtain different values of the distance between the solitons, $d_{sol}$,
by choosing appropriate values for
$H$ (and hence $k$) as described in appendix A.

We first show that the most significant contribution to long-range entanglement
originates from the solitons.
Consider a certain double-soliton configuration with solitons size $l_{sol}$ and separation $d_{sol}$.
Now let us take two blocks with the same size and separation: $l=l_{sol}$, $d=d_{sol}$ and
slide them along the chain.
We expect that as the blocks' positions become identical to the solitons' positions
we will get the maximal value of LN.
In figure \ref{sliding} we present $\log E_{LN}$ of two such sliding blocks,
where $g=3000$, $l_{sol}=56$ and $d_{sol}=281$.
The positions of the solitons are shown for reference.
The $x$ coordinate in the graph $n$ is the average of the
centers of the two blocks $A$ and $B$.
Since the double-soliton solution is symmetric, the centers of the blocks coincide with the locations of the solitons
for $n=0$.
Indeed, the entanglement is maximal at that point.
In addition to the main maximum, there are secondary maxima at $n=\pm170$,
which are obtained when $n$ coincide with the center of the solitons.
These maxima correspond to the entanglement between the right and left tails of each of the solitons.
We note that in the vacuum sector two such sliding blocks
yield practically constant value of long-range entanglement.

The fact that long-range entanglement has its peak 
at the core of the solitons is also manifested in
the structure of the Williamson's modes,
the so-called participation functions \cite{modewise1},\cite{modewise2}:
\begin{equation}
z_j(n)=u_j(n) v_j(n),
\end{equation}
where $H_AG_A u_j=\lambda_j^2 u_j$ and $G_AH_A v_j=\lambda_j^2 v_j$.
The participation functions qualitatively describe the distribution of entanglement between two
complementary regions.
We split the chain into two equal-sized blocks.
In the vacuum sector the participation functions are practically "localized" \cite{modewise2}.
$z_1^{vac}$, which has the highest contribution to the entanglement,
originates at the boundary between the parts.
The contribution of the other modes ($z_2^{vac},z_3^{vac},\dots$) decreases exponentially.
The localization of $z_j^{vac}$ departs from the boundary as $j$ increases.
In contrast, in the double-soliton solution,
the participation functions are not localized and correspond to the entanglement between
the solitons.
In figure \ref{partic2} we present
$z_1^{sol}$, $z_2^{sol}$, $z_1^{vac}$ and $z_2^{vac}$ in logarithmic scale,
where $\eta_1$ (the first internal mode) is also plotted for reference.
In contrast to the participation functions
in the vacuum sector,
the participation functions in the solitons sector
peak at the core of the solitons.

\subsection{Long-range entanglement vs. the separation}
\begin{figure}[ht]
\center{
\includegraphics[width=3in]{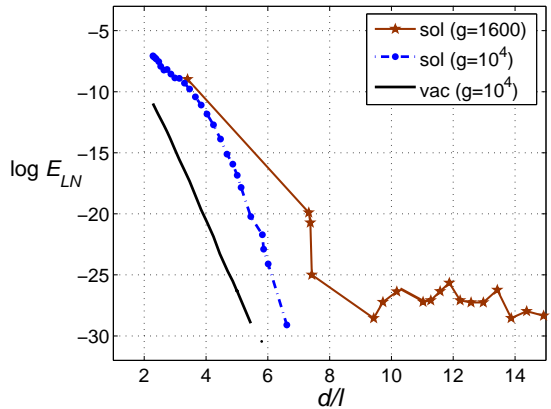}
\caption{(color online). Long range entanglement as a function of the solitons separation.
The centers of the two blocks coincide with the solitons:
$d=d_{sol}$ and $l=l_{sol}$.
The plots are given for $g=1600,10^4$.
$\log E_{LN}(g=1000)$ of the vacuum sector is plotted for reference, where $\beta(g)=5.57$.
$l\sim\sqrt{g}=40,96$ respectively.
In the case $g=10^4$, the entanglement of solitons decreases slower than
its critical vacuum sector correspondence
for small values of $d/l$.
In the case $g=1600$, the entanglement oscillates and is not a decreasing monotonic function
of the separation for large values of $d/l$.
}
\label{long_1}
}
\end{figure}
We now look at scaling of long-range entanglement as a function of the separation
between the solitons.
We compare long-range soliton-soliton entanglement to the
corresponding entanglement in the vacuum sector,
which is equivalent to
long-range entanglement in the linear harmonic chain.
In the {\it{critical}} vacuum sector
the scaling of the LN for
sufficiently large chains is universal and depends on a single parameter $r\equiv d/l$.
For sufficiently large values of $r$, the logarithmic negativity scales as $E_{LN}\approx e^{-\beta_c r}$,
where $\beta_c\approx2.7$ \cite{ln_hc} is the universal decay coefficient of the critical chain.
In the noncritical chain the logarithmic negativity is not universal.
Given two blocks with constant length $l$,
their entanglement scales as $E_{LN}\approx e^{-\beta(m,l)d/l}$
for $d>>l$, where $\beta(m,l)>\beta_c$ and $m$ is the mass of the particles.

In the soliton sector we find unique behavior of long-range entanglement,
which is qualitatively different from that of the vacuum sector.
Here we compute entanglement between two blocks that coincide with the solitons ($d=d_{sol}$)
for different values of $d_{sol}$.
The size of the blocks $l$ is chosen so that it maximizes von-Neumann entropy $E_S$
(Eq. \ref{ent}).
As expected $l$ does not depend on the separation $d=d_{sol}$ and approximately,
$l\sim\sqrt{g}$.
In figure \ref{long_1} we plot $\log E_{LN}$ versus $d/l$ for $g=1600,10^4$.
Both values of $g$ describe the noncritical regime.
(There is no critical regime in which the soliton solutions have internal modes.)
For reference we also plot $\log E_{LN}(g=10^4)$ for the corresponding vacuum sector
with the same value of $l$.
Clearly, long-range entanglement in the soliton sector is larger
than that in the vacuum sector (this is true for all values of $g$).
Note also that the scaling of LN in the soliton sector differs qualitatively from
that in the vacuum sector as $\log E_{LN}$
is not linearly dependent on $d/l$.
For relatively small values of $d/l$ a linear curve approximation of 
$\log E_{LN}^{sol}(g=10^4)$
yields $\beta=2.3$, which is
even stronger than the universal scaling of the critical vacuum sector \cite{ln_hc}.
In this regime
the solitons have a large overlap,
corresponding to the negative square mass in the region of the solitons.
As $d/l$ increases the linear curve does not fit and has a somewhat oscillating behavior.
For large values of $d/l$ we see that the decay of the entanglement
is larger than the corresponding one in the vacuum sector.

For the smaller value of the coupling constant, $g=1600$, we observe even more surprising behavior.
The LN {\it{oscillates}}
around a somewhat constant value for $d/l_0>8$.
In this regime long-range entanglement loses one of its most fundamental characteristics:
it is {\it{no longer}} a monotonically decreasing function of the distance between the solitons!

\subsection{Toy model for long-range entanglement}
\begin{figure}[ht]
\center{
\includegraphics[width=3in]{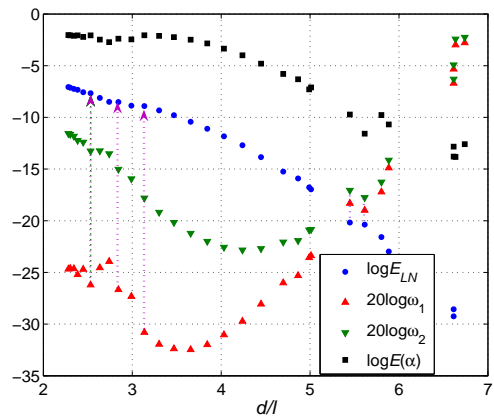}
\caption{(color online). Particle-like entanglement of the solitons:
dependence of $\log E_{LN}$ on the ratio between the internal modes' eigenfrequencies for
$g=10^4$, where $20\log\omega_1$ and $20\log\omega_2$
are also plotted.
The three left vertical arrows are examples of drastic increase of the gap,
which correspondingly changes
the entanglement.
Notice that as the gap is closed ($d/l>6$), the entanglement decreases significantly.
The entanglement of the toy-model, $E(\alpha)$, is shown for reference.
}
\label{particle_like}
}
\end{figure}

\begin{figure}[ht]
\center{
\includegraphics[width=3in]{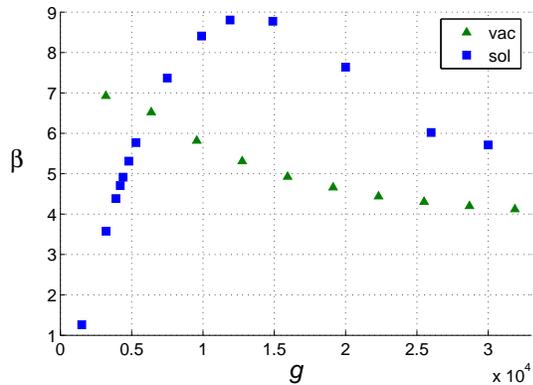}
\caption{(color online).
The exponential coefficient $\beta$ in the asymptotic behavior
$E_{LN}(d>>l)\sim e^{-\beta d/l}$
as a function of $g$.
In the vacuum sector $\beta$ decreases monotonically to the universal value $2.7$ (not seen in graph).
The lower bound $\beta>2$ is satisfied in the soliton sector for approximately $g>2000$.
For smaller values of $g$ the exponential approximation breaks down.
}
\label{long_2}
}
\end{figure}

We would like to analyze the unique behavior of the long-range entanglement in the soliton sector.
We discuss two factors, each of which dominates in a different regime of the coupling constant $g$.
For relatively large coupling, $g>5000$, the particle-like behavior of the
solitons is the dominant factor.
For relatively low coupling, $g<1000$, it is the topological nature of the first internal mode
that dictates the entanglement behavior.

Let us first discuss the particle-like behavior of the solitons by considering a toy-model.
From figure \ref{fig3} we see that the first two internal modes: $\eta_1$, $\eta_2$
are localized to the solitons
and describe qualitatively two groups of oscillating particles inside each of the solitons.
$\eta_1$ and $\eta_2$ are symmetric/antisymmetric and describes correlated/anticorrelated
oscillations of the two groups respectively.
Therefore, we simplify our model to
two coupled harmonic oscillators with normal-mode eigenfrequencies
$\omega_1$ and $\omega_2$,
which we identify with the internal modes' eigenfrequencies in the solitons sector.
Then it can be easily shown that the entanglement between the oscillators
depends on a single symplectic eigenvalue
$$\lambda=\frac{1}{4}\sqrt{2+\alpha+\alpha^{-1}},$$
where $\alpha=\omega_2/\omega_1$.
Correspondingly, the entanglement between the solitons
depends mainly on the ratio between the two internal modes eigenfrequencies.

In figure \ref{particle_like} we show the particle-like behavior of the two solitons
for $g=10^4$.
First note that both $\omega_1(d/l)$ and $\omega_2(d/l)$ are not monotonic functions.
Each has a certain global minimum, which appears due to the finiteness of the system.
As the gap between the eigenfrequencies is closed ($d/l>6$),
the entanglement decreases significantly.
Moreover, the slight deviations from the curve appear
when there are also deviations in the gap, as indicated by the vertical arrows.
For reference, we present the entanglement of our toy mode $E(\alpha)$.
Clearly, the toy-model manifests much larger entanglement
since the phonon band is not included,
and therefore its decohering screening effect is missing.

This model can be used to obtain a lower bound for the decay coefficient $\beta$ in the limit where $d>>l$.
In this limit we can use a double-well WKB approximation
to estimate the eigenfrequencies:
\begin{equation}
\frac{\Delta\omega}{\omega}\propto \exp\left({-\int\limits_{-d_{sol}/2}^{d_{sol}/2}|p(x)|dx}\right)=e^{-d_{sol}/\sqrt{g}},
\end{equation}
where $p(x)$ is the classical momentum, which is approximately $1/\sqrt{g}$.
In this case the entanglement can be approximated for $d>>l$ as
$E_S\propto e^{-2d/\sqrt{g}}$.
Therefore, we obtain a lower bound for $\beta$ in the soliton sector, $\beta\geq2$.
We expect that this bound is valid for large values of $g$, where the particle-like
behavior of the solitons is the most significant effect,
and for large values of $d/l$ so that the solitons have a small overlap.
However, for large values of the coupling constant
we cannot study very large values of $d/l$
since the system is finite.

Figure \ref{long_2}
shows the exponential decay coefficient in $E_{LN}^{sol}\sim e^{-\beta d/l}$
in the asymptotic limit $d>>l$, where the exponential approximation becomes accurate.
$\beta(E_{LN}^{vac})$ is also shown for reference.
In the vacuum sector $\beta$ decreases monotonically down to the
universal value $2.7$ (not seen in graph).
For $g>10^4$ the double-soliton sector has similar behavior.
However, for small values of $g$ the discussed lower bound is not satisfied, $\beta<2$,
as we enter the regime where $\omega_1$ is the most significant factor
and the exponential approximation breaks down.

\subsection{The noncritical regime}
\begin{figure}[ht]
\center{
\includegraphics[width=3in]{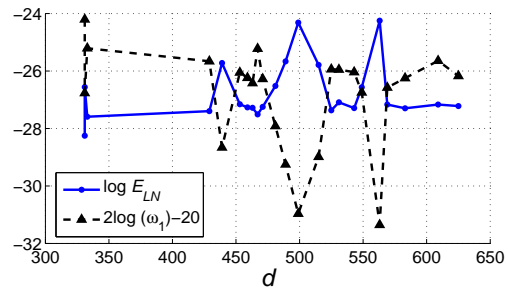}
\caption{(color online). Topological nature of $\omega_1$. $\log E_{LN}$ versus $d$ for $g=731$
(deep in the noncritical regime).
The computed logarithmic negativity is not a monotonically
decreasing function of the separation between the solitons.
For reference we plot $2\log(\omega_1)-20$.
Note that maximal entanglement is achieved for minimal $w_1$, which characterizes
the regime in which $g$ is small.
}
\label{delicate}
}
\end{figure}

\begin{figure}[ht]
\center{
\includegraphics[width=3in]{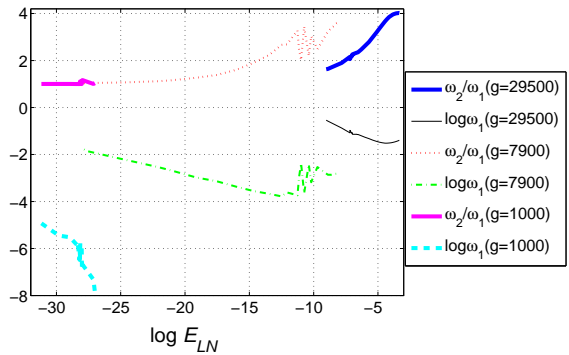}
\caption{(color online). $\omega_2/\omega_1$ and $\log\omega_1$
as a function of $\log E_{LN}$ for three coupling-strengths: $g=29500,7900,1000$.
}
\label{regimes}
}
\end{figure}

Let us now turn to the second factor, which dominates for relatively low coupling ($g<1000$).
As described in Appendix A the double-soliton solutions depart from
$\phi_{SG}$ (Eq. \ref{sgkink}) as we require a finite system and impose
$d\phi(\pm L)/dx\neq0$.
The separation of the solitons $d_{sol}$ is given by a parameter $k$,
where $d_{sol}$ monotonically increases with $k$.
Stability ($w_l^2\geq0$) regions of the system are defined in separate intervals of $k$.
As the values of $k$ in an arbitrary interval get closer to the interval boundaries,
$\omega_1$ becomes closer to zero.
Therefore, we obtain an oscillatory behavior of $\omega_1$ with the separation $d_{sol}$,
which strengthens as discreteness increases ($g$ decreases).
In figure \ref{delicate} we present $\log E_{LN}$ as a function of the separation between the solitons
$d_{sol}$,
where $g=731$.
For reference we present on the same graph $2\log(\omega_1)-20$ so that both plots are
more or less on the same magnitude.
Note that maximal entanglement is achieved for minimal $w_1$.
Clearly, in this regime long-range entanglement
is {\it{no longer}} a monotonically decreasing function of the distance between the solitons.

In figure \ref{regimes} we show both $\omega_2/\omega_1$ and $\log\omega_1$
as a function of $\log E_{LN}$ for three coupling values $g=29500,7900,1000$.
In the close to critical regime ($g=29500$) the particle-like behavior of the solitons is
a dominant factor as it is a monotonic function of the entanglement,
where the first internal mode has no significant effect
(its minimal value does not correspond to maximal entanglement).
In the deep noncritical regime the particle like behavior has no significant effect
as $\omega_2/\omega_1$ is approximately constant (slightly above $1$),
and the dominant factor is $w_1$.
In the case $g=7900$, we see that in general,
the entanglement increases with $\omega_2/\omega_1$ and
decreases with $\omega_1$.
However, there is an intermediate region,
where the entanglement is not a monotonic function of both factors.

\section{The weak-coupling limit}

\begin{figure}[ht]
\center{
\includegraphics[width=3in]{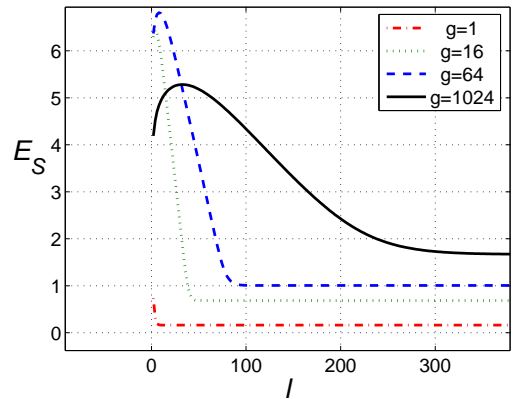}
\caption{(color online). $E_{S}$ as a function of $l$ for small values of $g$.
}
\label{discrete}
}
\end{figure}

We turn now to the weak-coupling limit of the FK model,
which is characterized by small values of $g$.
Here the entanglement has far less dramatic effects.
No long-range entanglement has been detected between two kinks.
In addition, the von Neumann entropy of an increasing block inside the kink loses its localization
as the coupling constant becomes too low, as seen in figure \ref{discrete}.
For $0.25\geq g\leq2.25$ there are two internal modes
in the soliton sector of the FK model \cite{kivshar2},
where the second internal mode has an eigenfrequency close to one.
However, the additional internal mode has no significant contribution to the entanglement of the kinks.

\section{Solitons as carriers of quantum information }

In this section we suggest and investigate the possible use of
solitons as carriers of quantum information.
We observed in the previous
sections that the entanglement properties of solitons in their ground state
are strongly related to the presence of internal localized modes.
These are (usually) the lowest discrete modes in the spectrum of the normal modes.
The internal modes qualitatively describe collective
or vibrational perturbations of the classical soliton.
In our approximation method the soliton is handled as a non-perturbative
classical object while the internal modes are linear perturbations around it.
The quantized internal modes then correspond to "phonon" bound states.

Our suggestion therefore is to use the localized internal modes' degrees of freedom
as carriers of quantum information.
Since the modes are "attached" to the soliton,
by moving the soliton under suitable adiabatic conditions,
information may be transported without being disturbed or mixed with the phonon band.
It is also possible to formulate multi-soliton processes
which create entanglement, and therefore
manipulate the information carried by each of the solitons.

To our knowledge this idea has not been investigated.
Therefore we provide a preliminary study of this possibility
without elaborating on a particular implementation.
In our model information will be coded by exciting the internal mode(s).
A fundamental requirement in quantum information processing
is the possibility of local operations.
However, the internal modes are collective degrees of freedom of all the particles in the system.
To resolve this difficulty one has to show that a local addressing of
the particles in the core of the soliton is sufficient in order to
manipulate the internal mode(s) with high fidelity.
Such a process seems possible
since the internal modes are localized.

To support the above requirement
we present two results regarding the single soliton and double-soliton respectively:

(1) Suppose we perform two-mode-squeezing of the internal mode
with some arbitrary external degrees of freedom, $Q$.
Then the entanglement inserted by squeezing
is saturated by the entanglement of local degrees of freedom
inside the soliton.

(2) Suppose we perform two-mode squeezing of two new collective modes
that are linear combinations
of the internal modes,
such that the new modes are confined to the left and right
sides of the chain respectively.
Then the entanglement inserted by squeezing is
saturated by local degrees of freedom
inside each of the two solitons.

Both of these results are crucial for any quantum information
applications, as otherwise operations cannot be implemented locally.

\subsection{Single soliton}
\begin{figure}[ht]
\center{
\includegraphics[width=3in]{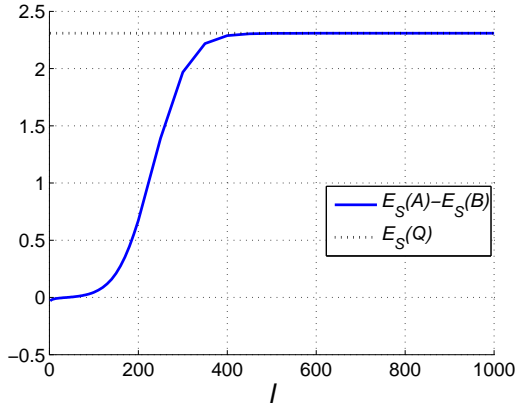}
\caption{(color online). Entangling a soliton and an external oscillator Q by
two-mode squeezing with $r=0.99$.
Shown $E_{S}(A)-E_{S}(B)\leq E_D(A)$ as a function of $l$ for $g=1000$.
$E_S(Q)$ is shown for reference.
Note that the distillable entanglement saturates the inserted entanglement
for quite small values of $l$.
}
\label{sqQ}
}
\end{figure}

Let us start with the first requirement.
We write the normal modes of the system as:
\begin{equation}
\begin{split}
&\tilde\phi_l=\sqrt{\frac{1}{2\omega_l}}\left(a_l e^{i\omega_l t}+a_l^\dagger e^{-i \omega_l t}\right),
\\&\tilde\pi_l=-i\sqrt{\frac{\omega_l}{2}}\left(a_l e^{i\omega_l t}-a_l^\dagger e^{-i \omega_l t}\right),
\end{split}
\end{equation}
where Eq. \ref{fields} can be rewritten as $\phi=\eta \tilde\phi$ and $\pi=\eta \tilde\pi$ in matrix form.
The degrees of freedom of the internal mode are $(\tilde\phi_1,\tilde\pi_1)$.
The ground state of the system is clearly
\begin{equation}
|\tilde\psi_{l=1,2,\dots,N}\rangle=\prod_{l=1}^N|\tilde 0_l\rangle\otimes|0_Q\rangle.
\end{equation}
Two-mode squeezing $S(r)$ of the internal mode and $Q$ is defined by
\begin{equation}
S(r) =e^{r(a_{1}^\dagger a_{Q}^\dagger-a_{1} a_{Q})},
\end{equation}
where $r$ is the squeezing parameter.
The squeezing acts on the internal mode and $Q$ directly through:
\begin{equation}
\label{sq1xp}
\begin{split}
& \tilde\phi_1\rightarrow(e^{+r}\tilde\phi_1+e^{-r}x_Q)/\sqrt{2},
\\& \tilde\pi_1\rightarrow(e^{-r}\tilde\pi_1+e^{+r}p_Q)/\sqrt{2},
\\& x_Q\rightarrow(e^{+r}\tilde\phi_1-e^{-r}x_Q)/\sqrt{2},
\\& p_Q\rightarrow(e^{-r}\tilde\pi_1-e^{+r}p_Q)/\sqrt{2},
\end{split}
\end{equation}
We can plug $S(r)$ into the covariance matrix
\begin{equation}
M\rightarrow\rm{Re}\left(\eta_2^{-1}S \tilde M S^{-1} \eta_2\right),
\end{equation}
where $\eta_2=\eta\oplus1\oplus\eta\oplus1$
and
$$\tilde M=\frac{1}{2}\rm{Diag}\left(\frac{1}{\omega_1},\frac{1}{\omega_2},\dots,\frac{1}{\omega_N},
\frac{1}\omega_Q,\omega_1,\omega_2,\dots,\omega_N,\omega_Q\right)$$
is the diagonal vacuum covariance matrix in the normal modes.

\begin{figure}[ht]
\center{
\includegraphics[width=3in]{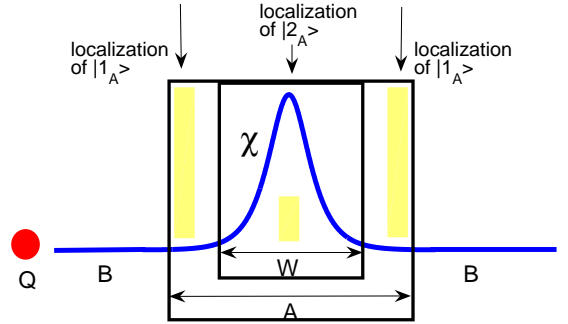}
\caption{(color online). Saturation of the hashing inequality -- schematic presentation.
At $t=0$ we perform two-mode squeezing of $\chi$ and $Q$ so that $\chi$ decay
significantly outside the squeezing block $W$ and
we measure the entanglement between $A$ and $Q$ such that $A>>W$.
In a simplified mode-wise decomposition of $A$ and $B\cup Q$ we treat only two modes:
$|1_A,1_B\rangle$, in which the participation function is localized at the boundary and
$|2_A,2_B\rangle$, in which the participation function is localized inside $W$.
}
\label{hash_fig}
}
\end{figure}

We show that for sufficiently large block $A$,
which includes the soliton,
the distillable entanglement $E_D$ between
local degrees of freedom inside $A$ and $Q$ saturates
the inserted entanglement via squeezing:
\begin{equation}
E_{S}=\cosh^2r \ln(\cosh^2 r)-\sinh^2r \ln(\sinh^2 r).
\label{inserted_ent}
\end{equation}
From hashing inequality \cite{plenio2},
\begin{equation}
E_D(A,Q)\!\geq\!\rm{max}\{0,\!E_S(A)\!-\!E_S(A\!\cup\! B),E_S(\!A\!)\!-\!E_S(\!A\!\cup \!Q\!)\},
\label{hashing}
\end{equation}
where $B$ complements $A$ in the chain so that
\begin{equation}
\begin{split}
&E_S(A\cup B)=E_S(Q),
\\&E_S(A\cup Q)=E_S(B).
\end{split}
\end{equation}
In figure \ref{sqQ} we plot $E_S(A)-E_S(B)$ as a function of $l$ for $g=1000$.
$E_S(Q)$ is shown for reference.
Note that indeed $E_D(A,Q)$ saturates $E_S(Q)$ for $l\geq450$.


We would like to provide a qualitative explanation
of the saturation obtained in figure \ref{sqQ}.
Note that the discussed saturation
does not depend on the existence of an internal mode,
but on the shape of the squeezing.
The internal modes are required for the entanglement to stay in the squeezing window,
but at $t=0$ they do not have a significant role, where an arbitrary collective mode $\chi$,
which has the shape of a bound state,
will be saturated by local modes.
We therefore turn to analyze the vacuum sector.
We specify two blocks in the chain as can be seen in figure \ref{hash_fig}:
the squeezing block $W$ and the entanglement measuring block $A$, where $A>>W$.
We measure the entanglement between $Q$ and $A$.
In addition, let us assume that $\chi$ decays significantly outside $W$.
Saturation of the hashing inequality means:
$E_S(A)-E_S(B)\rightarrow E_S(Q)$.

As the whole state is Gaussian we use the mode-wise decomposition \cite{modewise2,modewise1}.
That is, the state is described by a product of composed collective
modes from $A$ and $B\cup Q$:
\begin{equation}
\label{purewill}
|\psi\rangle=\prod|\psi_i\rangle_{ A,  (B\cup Q)}.
\end{equation}
Now before squeezing, the most significant contributions to
$E_S(A)(=E_S(B))$ come from collective modes that are localized
at the boundaries between $A$ and $B$.
The localization of less significant modes
tends to the center of $A$.
Therefore, modes that are well concentrated inside $W$
contribute very little to the initial entanglement.
The hashing inequality is saturated if after squeezing, $E_S(B)$ does not change much,
that is, almost all entanglement that is inserted by squeezing is added to $E_S(A)$.

To see that this is the case consider the following simplified model.
Let us assume that before squeezing the state can be expressed as a product of two composite states:
\begin{equation}
|\psi\rangle=|1_A,1_B\rangle\otimes|2_A,2_B\rangle
\end{equation}
where $|1_A,1_B\rangle$ is a strongly entangled mode, such that $|1_A\rangle$ is localized
at the boundaries of $A$ (outside $W$)
and $|2_A,2_B\rangle$ is a weakly entangled mode, such that $|2_A\rangle$ is localized
inside $W$.
Squeezing entangles $|2_A\rangle$ to $Q$ but it only weakly entangles  $|1_A\rangle$ to $Q$.
The entanglement between $B$ and $Q$ is therefore modified by two negligible contributions:
one from weak squeezing ($1_B$)
and the second from weak entanglement in the first place ($2_B$).

\subsection{Double-soliton}
\begin{figure}[ht]
\center{
\includegraphics[width=3in]{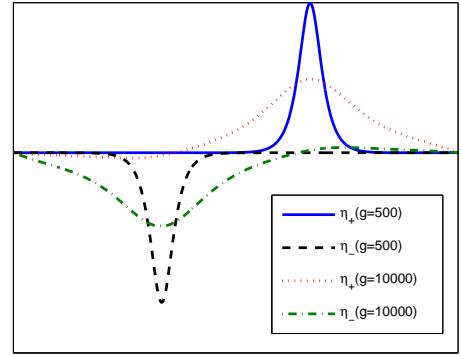}
\caption{(color online). Double soliton solution. $\eta_{\pm}=\frac{1}{\sqrt{2}}(\eta_1\pm\eta_2)$ for $g=500$
and $g=10^4$, where $\omega_1(g=500)=0.0011$, 
$\omega_2(g=500)=0.0012$, $\omega_1(g=10000)=0.0832$ and
$\omega_2(g=10000)=0.3075$.
Note that for almost degenerate modes the transformed collective modes $\eta_{\pm}$
are well separated,
while for nondegenerate modes the collective modes overlap and the squeezing is not optimal.
}
\label{collective}
}
\end{figure}

\begin{figure}[ht]
\center{
\includegraphics[width=3in]{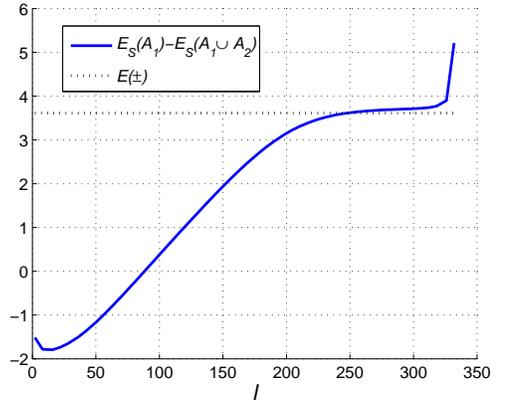}
\caption{(color online). Double soliton solution. Two-mode squeezing of $\eta_{\pm}$
with $r=2$.
The graph shows $E_{S}(A_1)-E_{S}(A_1\cup A_2)\leq E_D(A_1,A_2)$ as a function of $l$ for
$g=500$, $k=0.9415$ and $d(l)=331-l$
(the separation between the blocks decreases as their size increases).
$E(\pm)$ is plotted for reference.
Note that the distillable entanglement actually exceeds the inserted entanglement by squeezing,
as for large values of $l$, $d(l)$ decreases to very small numbers,
for which the vacuum entanglement has a non-negligible contribution.
}
\label{sqQ2}
}
\end{figure}

We turn now to the second result regarding double-soliton solutions.
Now the two internal modes are symmetric and antisymmetric respectively.
They are not confined to a single soliton each.
We want to create entanglement between the solitons by squeezing.
Therefore, we should use collective modes, where each is confined to a single soliton.
Let us define the linear transformation:
\begin{equation}
\eta_{\pm}=\frac{1}{\sqrt{2}}(\eta_1\pm\eta_2).
\end{equation}
In the case the internal modes are almost degenerate,
$\eta_{\pm}$ are localized well to each of the solitons.
Note that this condition corresponds to poor long-range {\it{vacuum}} entanglement of the solitons,
as described in Section $4$.
Practically degenerate internal modes appear only deep
in the noncritical regime of the double-soliton sector ($g<1000$),
as shown in figure \ref{collective},
where we plot $\eta_{\pm}$ for $g=500$ and $g=10^4$.

We therefore assume that $g=500$.
In correspondence with the single soliton squeezing, we define
\begin{equation}
S(r) =e^{r(a_{+}^\dagger a_{-}^\dagger-a_{+} a_{-})}
\end{equation}
as optimal squeezing of the two solitons.
We would like to show that the inserted entanglement by squeezing $E(\pm)$ (Eq. \ref{inserted_ent})
is saturated by local degrees of freedom
inside the two solitons: $A_1$ and $A_2$ with the same size.
From symmetry, $E_S(A_1)=E_S(A_2)$.
Hashing inequality implies that,
\begin{equation}
E_D(A_1,A_2)\geq E_S(A_1)-E_S(A_1 \cup A_2).
\label{hashing2}
\end{equation}
In figure \ref{sqQ2} we plot $E_S(A_1)-E_S(A_1 \cup A_2)$ as a function of blocks size $l$,
where the center of the blocks coincide with the center of the solitons.
$E(\pm)$ is plotted for reference.
Note that the distillable entanglement saturates the inserted entanglement for
$l\geq 250$.
Therefore, local degrees of freedom saturate $E(\pm)$.
Interestingly, the distillable entanglement actually exceeds the inserted entanglement through squeezing.
This is explained by the fact that for large blocks the distance between them becomes very small,
such that the vacuum entanglement has a non-negligible contribution.

\subsection{Implementations}
Finally, based on the above analysis, we would like to briefly discuss two possible applications:
entanglement transportation and entanglement manipulation through
a tunneling gate.
In the entanglement transportation task 
we assume there is an additional
external degree of freedom, $Q$.
Initially the soliton is located in one side of the system, next to, say, Alice.
Alice entangles $Q$ to some {\it{local}} degrees of freedom inside the soliton.
Then she classically transport the soliton to the other side, next to Bob.
Finally, Bob possesses local degrees of freedom inside the soliton with which he can, {\it{e.g.}},
perform experiments that prove nonlocal correlations with $Q$.
In order to keep the state Gaussian, the entanglement is realized by two-mode squeezing.
In addition, we assume slow transportation so that the quantum modes are perturbed only adiabatically.

In the second implementation we use the double-soliton sector.
Quantum information is tunneled between the solitons
by bringing them close to each other for a finite time interval
and then separate them apart.
This application is much more complicated than the first one,
as we require that the phonon band is not excited during the tunneling phase,
while the internal modes do.

\section{Conclusions}
We conclude with the main results in this work
and  proposals for further study.
We have shown that the
entanglement of solitons manifests particle-like
behavior as they are characterized
by localization of entanglement and long-range entanglement.
The von-Neumann entropy of solitons mixes both critical and noncritical behaviors,
where it increases logarithmically inside the core of the soliton, reaches a maximal value and then
decreases and saturates the constant value that
corresponds to the massive vacuum sector.
Interestingly, the increase of the entanglement
inside the core is faster than that in a universal
critical field.
In addition, we have shown that two solitons manifest long-range entanglement,
which may not decrease with their separation in the noncritical regime.
Near the critical regime the entanglement
decreases slower than the corresponding universal
critical field.

Our quantum model is based on linear perturbations
around a nonlinear classic solution.
Higher order corrections \cite{tambolis} are important
for the treatment of the zero mode
in the infinite system.
In this case one obtains non-linear terms which couple
the solitons' collective coordinate with the phonon modes.
It would be interesting to explore classical
configurations of two non-static solitons
which can produce entanglement through their mutual phonon modes.
Moreover, certain models, such as the $\phi^4$,
manifest in addition to the zero mode
a nonzero internal mode which is associated with local shape deformations.
It would be interesting to study mechanisms
for entanglement of the internal modes in
soliton-soliton scattering \cite{campbell}
and in soliton-impurity scattering  \cite{kivshar_imp}.

We suggest that solitons' internal modes be used as carriers of quantum information.
Since the modes are "attached" to the soliton,
by moving the soliton under suitable adiabatic conditions,
information may be transported without being disturbed or mixed with the phonon band.
Our proposal differs from previously suggested methods,
which employ a collective displacement of the soliton
using long Josephson junctions \cite{ustinov}.
Realizations of the present suggestion can possibly be achieved
either in discrete systems such as trapped ions or in an effectively one-dimensional 
Bose-Einstein condensate, which manifests soliton solutions \cite{bec,bec2}.
Recently, a possible realization of an FK-like model in the ion trap was discussed
in \cite{fk_ion}. 
The utilization of internal modes in trapped ion systems will
be discussed in a future work \cite{hagai}.

\acknowledgments
We thank A. Retzker, H. Landa, J. Kupferman and M. Marcovitch for helpful discussions.
This work has been supported by the Israel science foundation grant
no. 784/06 and German-Israeli foundation Grant no. I-857.

\appendix\section{Solutions of the sine-Gordon equation for finite systems}
\renewcommand{\theequation}{\thesection.\arabic{equation}}
\setcounter{equation}{0}
For a comprehensive analysis see \cite{kuplevakhsky}.
The finite sine-Gordon system describes the magnetic flux in long Josephson junctions with constant,
homogeneous external magnetic field $H$ perpendicular to the barrier
and bias current line density $I$.
We use dimensionless units:
$$x\rightarrow x/\lambda_J,\quad H\rightarrow H/H_s,\quad I\rightarrow I/I_s,$$
where $\lambda_J$ is Josephson length and
$$H_s=I_s=\big[e(2\lambda_L+t)\lambda_J\big]^{-1}$$
is the superheating field of the
vortex-free Meissner state,
where $t$ is the thickness of the barrier and $\lambda_L$
is London penetration depth.

In the finite system the boundary condition we impose on the magnetic flux $\phi$ is:
\begin{equation}
\label{boundary}
\frac{d\phi}{dx}(\pm L)=\pm I+2H
\end{equation}
where $2L$ is the length of the barrier in Josephson's length units.
Imposing stationary condition on sine-Gordon equation, Eq. \ref{sg} reduces to
$$d^2\phi/d x^2=\sin\phi,$$
where for zero transport currents the solutions are:
\begin{equation}
\label{elliptic}
\begin{split}
&\phi_e(x)=\pi (\sigma-1)+2am(\frac{x}{k}+K(k),k),\quad \sigma=0,2,4,\dots,
\\&\phi_o(x)=\pi \sigma+2am(\frac{x}{k},k),\quad \sigma=1,3,5,\dots,
\end{split}
\end{equation}
where 
$am$ is the Jacobi elliptic amplitude and $K$ is the complete elliptic integral of the first kind.
The subscripts $e$ (even) and $o$ (odd) refer to the number of soliton (vortices) in the solutions.

These solutions are stable only for specific conditions on $k,L$ and $H$.
Stability regions in terms of $k$ and $H$ for Eq. \ref{elliptic} are established from the following:
first, the roots of the equation
$$\sigma K(k)=L,\quad \sigma=1,2,\dots,$$ form an infinite decreasing sequence of bifurcation points,
where $\sigma$ is the analog of the topological index $\sigma$ in the infinite single soliton sector.
$$k=k_\sigma\in I\equiv(0,1]=\cup_{\sigma=0}^{\infty}I_\sigma,$$
$$I_0=(k_1,1],$$
$$I_\sigma=(k_{\sigma+1},k_\sigma],$$
where $\sigma=1,2,\dots$
The stability regions for the solutions
$\phi=\phi_e$ are given by the intervals $I_{2m}$,
whereas the stability regions
for the solutions $\phi=\phi_o$ are given by the intervals
$I_{2m+1}$, where $m=0,1,2,\dots$
Dependence of $k$ on $H$ is given by
\begin{equation}
\begin{split}
&dn\Big(\frac{L}{k},k\Big)=\frac{1-k^2}{kH},\quad \sigma=2m,
\\&dn\Big(\frac{L}{k},k\Big)=kH,\quad \sigma=2m+1,
\end{split}
\end{equation}
where
$$\frac{\partial}{\partial u} dn(u,k)=-k \cdot cn(u,k) sn(u,k)$$
where $cn(u,k)$ and $sn(u,k)$ are correspondingly the elliptic cosine and sine functions.
The stability regions in terms of the field $H$ take the form:
\begin{equation}
\label{hstability}
\begin{split}
&0\leq H<H_0,\quad \sigma=0
\\&\sqrt{H_{\sigma-1}^2-1}\leq H<H_\sigma,\quad \sigma=1,2,\dots
\end{split}
\end{equation}
with $H_\sigma$ implicitly determined by
$$(\sigma+1)K\Big(\frac{1}{H_\sigma}\Big)=H_\sigma L.$$

Turning to the quantization of the perturbation,
one has to solve Eq. \ref{stability}:
\begin{equation}
\label{lame}
-\eta''+\left[2k sn^2(x+x_0,k)-1\right]\eta=\omega^2\eta
\end{equation}
where $x_0$ is the phase of the center of the junction.
Eq. \ref{lame} is also known as Lam\'{e} equation.
The boundary conditions for $\eta$ are
$\eta'|_{x=\pm L}=0$ \cite{remark_eta}.

Finally we would like to note that
if bias current were introduced,
the locations of the solitons in the presented solutions could
be modified to a nonsymmetric configuration.
In addition, note that $g$ does not appear in the equations.
Recall that in the infinite single soliton solution
the width of the soliton is $\sqrt{g}$.
In the continuous finite-system case one can not change arbitrarily the width of the vortices,
as they scale proportionally with the system size.
Throughout the analysis we choose $L=8$ as the sampled analytical solution we start with in the
minimizing algorithm.


\begin{thebibliography}{99}
\bibitem{remark_omega} For periodic boundary conditions
translational symmetry implies that the minimal
eigenfrequency of the phonon band is zero,
which results in an infinite contribution to the entanglement
This problem is absent in our analysis.
\bibitem{remark_N} Throughout the analysis we take $N=1000$ particles for the computation.
The results do not change qualitatively for larger values.
\bibitem{remark_alpha} the $\alpha$ in \cite{modewise2} is related to $g$ by $\alpha=2g/(2g+1)$.
\bibitem{remark_eta} In our analysis we chose $\eta(\pm L)=0$.
The change of the boundary conditions, however, has no effect on the entanglement.
\end{thebibliography}
\end{document}